\documentstyle[11pt,epsfig]{article} 
\def\baselinestretch{1.3}
\newcommand{\ba}{\begin{array}}
\newcommand{\ea}{\end{array}}
\newcommand{\bd}{\begin{displaymath}}
\newcommand{\ed}{\end{displaymath}}
\newcommand{\be}{\begin{equation}}
\newcommand{\ee}{\end{equation}}
\newcommand{\bea}{\begin{eqnarray}}
\newcommand{\eea}{\end{eqnarray}}

\def\e{\epsilon}

\def\l{\lambda}
\def\m{\mu}
\def\n{\nu}

\def\q2 {q^2}

\begin{document}
\begin{flushright}
{\large  MRI-PHY/P990513 \\May, 1999  \\ hep-ph/9905549}
\end{flushright}

\begin{center}
{\Large\bf Constraining an  $R$-parity violating supersymmetric theory 
from the SuperKamiokande data on atmospheric neutrinos }\\[20mm]
Aseshkrishna Datta \footnote{E-mail: asesh@mri.ernet.in}, 
Biswarup Mukhopadhyaya\footnote{E-mail: biswarup@mri.ernet.in}
and Sourov Roy \footnote{E-mail: sourov@mri.ernet.in} \\
{\em Mehta Research Institute,\\
Chhatnag Road, Jhusi, Allahabad - 211 019, India}, 
\\[20mm] 
\end{center}
\begin{abstract}
The constraints  on an R-parity violating supersymmetric theory   arising 
from the recent SuperKamiokande results on atmospheric neutrinos are studied,
with special reference to a scenario with bilinear R-parity violating terms. 
Considering both the fermionic and scalar sectors, we
find that a large area of the parameter space is allowed, in terms of both 
the lepton-number violating entries in the superpotential and the
soft R-violating terms in the scalar potential, and that no fine-tuning 
is required. However, the need to avoid flavour changing neutral currents
puts additional restrictions on the theory, requiring either the R-violating
terms in the superpotential to be smaller than the R-conserving ones, or
a hierarchy in the  R-violating parameters for different lepton 
flavours in the superpotential. 
    
\end{abstract}

\vskip 1 true cm

\newpage
\setcounter{footnote}{0}

\def\baselinestretch{1.8}

\section{Introduction}

Ever since the evidence in favour of neutrino oscillation has received
reinforcement from the SuperKamiokande(SK) results on the atmospheric
$\nu_\mu$-defficiency \cite{SK},  intensive discussions have taken place on 
 `new physics' that can give rise to neutrino masses
and mixing of the expected types. In the simplest explanation,
the defficiency is due to the oscillation of $\nu_\mu$ into $\nu_\tau$,
with $\Delta m^2 \simeq 10^{-3}$ eV $^{2}$ and large angle vaccum mixing
($\sin^2 2\theta > 0.8$) between the two neutrino species. Such oscillation
is also consistent with the results from the SOUDAN \cite{SOU} and MACRO
\cite{MAC} experiments.
Side by side, one
also faces the requirement of accounting for the solar neutrino deficit
\cite{SOL}.
Using the still available explanations in terms of  $\nu_e - \nu_\mu$ 
oscillation, this
means a mass-squared difference of about $10^{-5} - 10^{-6}$  eV$^2$ for
the MSW solution (with either small or large angle mixing), or a mass-squared
splitting smaller by about 4-5 orders for the vacuum oscillation solution.
Thus one faces the task of reconciling a mass hierarchy with 
large angle mixing between the two heaviest neutrino species.  In whichever
way that is possible, one has to step outside the domain of the standard
electroweak model.

Supersymmetry (SUSY) has been studied for a long time now, from both
theoretical and phenomenological angles, as one of the most attractive
options beyond the standard model \cite{SUSY}. And yet, apart from indirect  
successes such as offering solutions to the hierarchy puzzle,
it has not been possible to confront SUSY with any clear experimental
results so far. It is therefore appropriate that when the atmospheric
neutrino results are so emphatically underlining the existence of neutrino 
masses and mixing with a given pattern, the relevance of SUSY in generating
and explaining such a pattern should be thoroughly explored.   
 
If the SUSY extension of the standard model (SM) with the minimal particle 
content has to be invoked for the purpose, then the absence of any
right-handed neutrino (and therefore of Dirac masses) in the 
scenario immediately points towards Majorana neutrinos as the likely
solution. However, the latter implies the violation of lepton number by 
2 units. That again can come rather naturally in a scenario where 
R-parity, defined as $R~~=~~(-1)^{3{\mathcal{B}+\mathcal{L}}+2{\mathcal S}}$, is violated 
\cite{RP}. In such a
scenario, the $\Delta {\mathcal{L}} = 1$ terms in the Lagrangian can ultimately give rise
to Majorana masses either through a tree-level see-saw type mechanism or
via loop effects. The `attractive' point here is that one does not have to
postulate the existence of any particle {\it specifically} for the
generation of neutrino masses, since the superparticles that are 
indispensable components of a SUSY theory are sufficient for the purpose.

Perhaps the most convenient and cogent (though not unique) way of introducing
R-parity violation is an extension of the superpotential, using
trinilear and/or bilinear terms. Trilinear terms 
in the superpotential \cite{TRI} give rise to $\mathcal{L}$ 
(or $\mathcal{B}$)-violating 
Yukawa-type 
interactions and trilinear soft terms in the scalar potential. 
The presence of bilinear $\mathcal{L}$-violating terms 
\cite{BILI1, BILI2, BILI3}, 
on the other hand, 
are generally responsible for non-vanishing vacuum expectation values 
(vev) for sneutrinos, which also lead to mixing between neutrinos and
neutralinos as also between charged leptons and charginos (and similarly
between the Higgs and charged slepton/sneutrino states in the scalar sector).
The bearing  of both approaches on neutrino masses  have been studied 
extensively in recent times, whereby  constraints on the R-parity violating 
parameters from neutrino masses have been derived \cite{Triset1,
Bilset1}. One also finds in the 
literature discussions on how to test the consequences
of the corresponding scenarios in accelerator-based experiments
\cite{BILEX}. 

In several of these references,  it was shown how one could accommodate 
the mass hierarchy together with large angle mixing between  $\nu_\mu$ and 
$\nu_\tau$ through neutrino-neutralino mixing via the bilinear terms mentioned
above, while the smaller mass splitting between the   $\nu_\mu$ and $\nu_e$
could be due to loop-induced  effects. However, one can still  
ask a number of questions related to the parameter space of the
theory, extending both over the fermionic and the scalar sector, 
before the SUSY explanation of the SK results can acquire 
sufficient 
credibility. In this paper we have tried to find answers to some such 
questions, and to establish that the  solution space for the 
SK deficits is {\em not} a fine-tuned one.

To be more specific, our analysis includes both the scalar and spin-1/2 
sectors of an R-parity violating
scenario (where the simplest, bilinear R-violating terms are introduced as
necessary ingredients but no generality is otherwise discarded) which
can explain the SK data. Keeping the value of 
$\Delta m^2_{\mu\tau}$ and the large mixing angle as inputs, we have tried 
to find as general answers as possible to the following questions:

\begin{itemize}  
\item Is it a necessity to have a large hierarchy between the
      bilinear terms corresponding to different lepton families in the 
      superpotential?

\item Is it required to have a hierarchy between soft $\mathcal{L}$-violating 
terms involving different lepton flavours?

\item How crucial is it to have a hierarchy of orders between the
      $\mathcal{L}$-conserving and $\mathcal{L}$-violating bilinear terms in 
the superpotential and the scalar potential?

\item How is the suppression of flavour changing neutral currents (FCNC)
      ensured in such a picture?   
 
\end{itemize}

In this analysis, we have used low-energy values of all the parameters 
in the theory, and have not attempted to link them to any specific 
high-scale physics. However, our chosen convention of writing the soft
bilinear terms in the scalar potential carries some influence of a
supergravity (SUGRA)-based model where such terms can be shown to originate 
from an interference of terms belonging to the hidden and observable 
sectors in the superpotential.

Section 2 sets up the general framework within which we operate, 
incorporating the fermion and
scalar mixing schemes. The different choices of basis in which we have worked 
are specified there, and the parameters which we ultimately use to explore
large angle mixing in the most general sense are defined. Section
3 contains numerical studies of the constraints on the theory 
in terms of these parameters. This gives us full or partial answers to 
the questions listed above,  obtained either from detailed calculations or 
from simple estimates of order. We conclude in section 4.

\section{The Framework}

The MSSM superpotential is given by (suppressing the $SU(2)$ indices)

\begin{equation}
W_{MSSM} = {\mu} {\hat H}_1 {\hat H}_2 + h_{ij}^l {\hat L}_i {\hat
H}_1 {\hat E}_j^c
+ h_{ij}^d {\hat Q}_i {\hat H}_1 {\hat D}_j^c + h_{ij}^u {\hat Q}_i
{\hat H}_2 {\hat U}_j^c
\end{equation}
where $\m$ is the Higgsino mass parameter and the last three terms give all the 
Yukawa interactions.

When R-parity is violated, the following additional terms can be 
added to the superpotential:

\begin{equation}
W_{\not R} = \lambda_{ijk} {\hat L}_i {\hat L}_j {\hat E}_k^c +
\lambda_{ijk}' {\hat L}_i {\hat Q}_j {\hat D}_k^c +
\lambda_{ijk}''{\hat U}_i^c {\hat D}_j^c {\hat D}_k^c + \epsilon_i {\hat
L}_i {\hat H}_2
\end{equation}

Where the $\lambda''$-terms correspond to $\mathcal{B}$-violation, 
and the remaining
ones, to $\mathcal{L}$-violation. The absence of proton decay makes 
it customary to
have one of these two types of nonconservation at a time. In the rest of
this discussion, we shall not consider $\mathcal{B}$-violation.

The $\lambda$-and $\lambda'$-type terms have been widely studied;
their contributions to neutrino masses can be only through loops, and
their multitude (there are 36 of them altogether) makes the necessary
adjustments possible for creating the required values of neutrino masses
and mixing angles. 

More interesting, however, are the three bilinear
terms ${\epsilon}_{i}L_{i}H_{2}$. It is in fact quite useful to 
use them as the starting inputs of R-parity violation in the theory. 
There being at most only three parameters of this type, the model looks
much more predictive than one with 36 unrelated trilinear terms. 
Furthermore, the physical effects of the trilinear terms
can be generated from the bilinears, by going to the appropriate bases.
There are additional interesting consequences of the bilinear terms.
The presence of the $LH_2$-term means a mixing between the Higgsinos and
charged and neutral lepton states. In addition, the scalar potential in
such a case contains terms bilinear in the slepton and the higgs fields and
involving only second and third generation of sleptons (the reason behind this
assumption will be
clarified as we proceed) the terms are as follows :

\begin{eqnarray}
V_{scal} = m^2_{L_3} {\tilde L}^2_3 ~+~ m^2_{L_2} {\tilde L}^2_2 ~+~ 
m^2_1 H^2_1 ~+~ m^2_2 H^2_2 ~+~ B \mu H_1 H_2 \nonumber
   \\[1.0ex]
~+~ B_2 \e_2 {\tilde L}_2 H_2 ~+~B_3 \e_3 {\tilde L}_3 H_2 
~+~ \m \e_3 {\tilde L}_3 H_1 ~+~ \m \e_2 {\tilde L}_2 H_1 ~+~ .....
\end{eqnarray}  

\noindent
where $m_{L_i}$ denotes the mass of the {\it i}-th scalar doublet, ${\tilde
L}_i = \left( \begin{array}{c}
{\tilde \n}_i \\
{\tilde l}_i
\end{array}  \right)_L$, {\it i} being 2 and 3 for the second and third
generations respectively, at the electroweak scale. Here $m_{L_2}$ 
and $m_{L_3}$ are the slepton mass parameters. In our subsequent analysis,
all the sleptons (of both chiralities) and sneutrinos have been assumed to
be degenerate.

An immediate consequence of the additional ($\mathcal{L}$-violating) soft terms in the
scalar potential is a set of non-vanishing sneutrino vev's \cite{SVEV}. 
This is a characteristic feature of this scenario, which in addition 
produces neutrino(charged lepton)-gaugino mixing via the 
sneutrino-neutrino(charged lepton)-gaugino interaction terms. The former
type of mixing leads to a neutrino mass at the tree-level.

Since our primary goal is to explain large angle $\nu_{\mu}-\nu_{\tau}$
mixing, we simplify the picture by assuming that only the second and third 
generations enter into this tree-level mixing process. For this, we postulate
only two bilinear R-violating terms, proportional to $\epsilon_2$ and
$\epsilon_3$. These two terms, together with the soft bilinear terms
$B_2$ and $B_3$, form the set of independent R-parity violating parameters
in this basis, henceforth to be called {\em basis 1}. 
The vev's corresponding to the muonic and tau sneutrinos are 
$v_{\mu}$ and $v_{\tau}$ respectively in this basis. For reasons that will
become apparent later, we choose to treat these vev's as independent
parameters, and use them to derive values of the soft terms. 
For that purpose, one has to make use of the set of tadpole equations
arising from electroweak symmetry breaking:
\begin{eqnarray}
(m^2_1 + 2 \lambda c)v_1 + B \mu v_2 + \mu \epsilon_2 v_\mu + \mu
\epsilon_3 v_\tau = 0 \\
(m^2_2 - 2 \lambda c)v_2 + B \mu v_1 + B_2 \epsilon_2 v_\mu + B_3
\epsilon_3 v_\tau = 0 \\
(m^2_{{\tilde \nu}_\mu} + 2 \lambda c)v_\mu + B_2 \epsilon_2 v_2 
+ \mu \epsilon_2 v_1 + \epsilon_2 \epsilon_3 v_\tau = 0 \\
(m^2_{{\tilde \nu}_\tau} + 2 \lambda c)v_\tau + B_3 \epsilon_3 v_2 
+ \mu \epsilon_3 v_1 + \epsilon_2 \epsilon_3 v_\mu = 0
\end{eqnarray}
where $v_1 = <H_1>$, $v_2 = <H_2>$, $c = (v_{1}^2 -v_{2}^2 
+ v_{\tau}^2 + v_{\m}^2)$ and $\l = (g^2 + {g'}^2)/8$.

While two of these equations can be used to eliminate the soft Higgs mass
terms $m_1$ and $m_2$, the ${\mathcal L}$-violating soft terms $B_2$ and 
$B_3$ can be obtained from the remaining two:
\begin{eqnarray}
B_2 = -{\frac {1} {\epsilon_2 v_2}} (\epsilon_2 \epsilon_3 v_\tau 
+ 2 \lambda c v_\mu + m^2_{{\tilde \nu}_\mu} v_\mu + \epsilon_2 \mu v_1) \\
B_3 = -{\frac {1} {\epsilon_3 v_2}} (\epsilon_2 \epsilon_3 v_\mu 
+ 2 \lambda c v_\tau + m^2_{{\tilde \nu}_\tau} v_\tau + \epsilon_3 \mu
v_1) 
\end{eqnarray}
\noindent
where, again, the sneutrino masses have been assumed to be degenerate with
a common slepton mass parameter.

The next step is to rotate away both the $\epsilon$-terms from the 
superpotential. In the process we go from the basis ($H_1$, $L_3$, $L_2$)
to ($H'_1$, $L'_3$, $L'_2$) using the following rotation :

\begin{equation}
\left( \begin{array}{c}
H'_1 \\
L'_3 \\
L'_2
\end{array}  \right) = 
\left( \begin{array}{ccc}
c_3 & s_3 c_2 & s_3 s_2 \\
-s_3 & c_3 c_2 & c_3 s_2 \\
0 & -s_2 & c_2 
\end{array}  \right)
\left( \begin{array}{c}
H_1 \\
L_3 \\
L_2
\end{array}  \right)  
\end{equation}

\noindent
where $s_2=\frac {\epsilon_2} {\sqrt{\epsilon^2_2 + \epsilon^2_3}}$,~~~ 
$c_2=\frac {\epsilon_3} {\sqrt{\epsilon^2_2 + \epsilon^2_3}}$,~~~
$c_3=\frac{\mu}{\mu^\prime}$, ~~~
$s_3={\frac{\sqrt{\epsilon^2_2 + \epsilon^2_3}}{\mu^{'}}}$, 
and $\m'=\sqrt{\mu^2 + \epsilon^2_2 + \epsilon^2_3}$.
Clearly, this leaves $\m' H'_1 H_2$ as the only bilinear term
in the superpotential. The physical consequences of bilinears R-parity
violation,
however, are still existent in this basis, since the scalar potential
contains its signature, and one has now `rotated' sneutrino vev's
$v'_\m$ and $v'_\tau$ which, together with the Higgs vev
$v'_1$, are connected to the set ($v_{1}, v_{\tau}, v_{\m}$) by the
rotation matrix given in Eqn. (10) above. 

In this basis (called {\em basis 2}), the neutralino mass matrix (which is of
dimension $6 \times 6$ after including the neutrinos) is given by

\begin{equation}
{\cal M}_\chi =  \left( \begin{array}{cccccc}
  0 & -\mu' & \frac {gv_2} {\sqrt{2}} &
  -\frac {g'v_2} {\sqrt{2}} & 0 & 0 \\
  -\mu' & 0 & -\frac {gv'_1} {\sqrt{2}}
       & \frac {g'v'_1} {\sqrt{2}} & 0 & 0 \\
 \frac {gv_2} {\sqrt{2}} & -\frac {gv'_1} {\sqrt{2}} & M & 0 & -\frac
 {gv'_\tau}
 {\sqrt{2}} & -\frac {gv'_\mu} {\sqrt{2}} \\
 -\frac {g'v_2} {\sqrt{2}} & \frac {g'v'_1} {\sqrt{2}} & 0 & M' &
  \frac {g'v'_\tau} {\sqrt {2}} & \frac {g'v'_\mu} {\sqrt {2}} \\
  0 & 0 & -\frac {gv'_\tau} {\sqrt {2}} & \frac {g'v'_\tau} {\sqrt {2}} &
  0 & 0  \\
  0 & 0 & -\frac {gv'_\mu} {\sqrt {2}} & \frac {g'v'_\mu} {\sqrt {2}} &
  0 & 0
  \end{array}  \right)
\end{equation}

\noindent
where the successive rows and columns correspond to
(${\tilde H}_2, {\tilde H'}_1, -i\tilde{W_3},
-i\tilde{B}, \nu'_\tau, \nu'_\mu$).
Here $M$ and $M'$ are the ${\rm SU(2)}$ and ${\rm U(1)}$
gaugino mass parameters respectively, $\mu'$ is 
the Higgsino mass parameter in basis {\it 2} and $v'_1 = <H'_1>$. 
This leads to {\em one} non-vanishing tree-level neutrino mass 
eigenvalue. The root of this can be traced to the fact that one linear 
combination of $\n'_\m$ and $\n'_\tau$, given by

\begin{equation}
\nu_3 = \n'_\tau \cos {\theta}   + \n'_\m \sin {\theta}
\end{equation}

\noindent
enters into cross-terms with the $\tilde{B}$ and $\tilde{W_3}$, while
its orthogonal state $\nu_2$ decouples from the mass matrix and
therefore remains massless \cite{polon}. The angle $\theta$ is given by 

\begin{equation}
\cos \theta~~= ~~ v'_\tau/v'
\end{equation}

\noindent
with $v'=~\sqrt{{v'}^2_\m +  {v'}^2_\tau}$. The quantity
$v'$,  
which is a kind of `effective' sneutrino vev in a basis where the bilinear
terms are rotated away from the superpotential, 
is a basis-independent
measure of R-parity violation, which directly  controls the tree-level 
neutrino mass acquired in the process. The mass implied by the
atmospheric $\nu_{\mu}$-deficit in the SK data requires $v'$ to be
in the range $(1-3) \times 10^{-4}$ GeV approximately \cite{viss}, 
depending on the mass of the lightest neutralino ($\chi^0_1$). It is also 
interesting to note 
at this stage that the smallness of the neutrino mass does not limit
the value of the $\epsilon$-parameters so long as   $v'$ lies within
about 100 keV or so. It has been demonstrated, for example,
that in theories based on $N$=1 supergravity (SUGRA), it is indeed possible to 
have a small $v'$ in spite of large values of the $\epsilon$'s, by
setting all bilinear soft terms (both $\mathcal{L}$-violating and conserving) 
to the same value at the SUGRA breaking scale \cite{sugra}.

Modulo the very small neutrino-neutralino mixing, the angle $\theta$ defined 
above can be identified with the neutrino vacuum mixing angle ($\theta_0$)
if the {\em charged lepton mass matrix is diagonal} in basis {\it 2}. Only 
in such a
case can we say that a near-equality of the vev's $v'_\m$ and
$v'_\tau$ is required by the condition of large-angle mixing. 
However, the general form of the charged lepton mass matrix in this basis 
is

\begin{equation}
{\cal{M}}_l  = \left( \begin{array}{cc}
f_3(-v'_\mu s_2 s_3 + c_2 v'_1) & 
f_2(s_2 v'_1 + c_2 s_3 v'_\mu) \\
f_3(-c_3s_2v'_1 + s_2 s_3 v'_\tau) &
f_2(c_2 c_3 v'_1 - c_2 s_3 v'_\tau) 
\end{array}  \right)
\end{equation}
Here we have assigned ($\tau'_L$, $\m'_L$) along the rows and 
($\tau_R$, $\m_R$) 
along the columns. One should note that $f_3 = h^l_{33}=\frac{m_{\tau}}{v_1}$ 
and $f_2 = h^l_{22}=\frac{m_\mu}{v_1}$.
Thus ${\mathcal{M}}_l$ has non-vanishing off-diagonal terms in general. 
The conditions under which 
it can be approximately treated as diagonal (and the equality of the
two sneutrino vev's in basis {\it 2} is a necessity) will be specified 
in the next section. Instead, we observe here that the actual neutrino 
mixing matrix is given by

\begin{equation}
V_{l} = V_{\theta} U^T
\end{equation}
 
\noindent
where $V_{\theta}$ is the $2 \times 2$ orthogonal matrix corresponding 
to the rotation angle $\theta$, 
and $U$ is the matrix that diagonalises ${\cal{M}}_l$ is basis {\it 2}. 
It is the matrix $V_l$ which should correspond to a rotation angle of
${\pi}/4$ for maximal mixing, and to a range approximately 
between 32 and 58 degrees
for $\sin^2 2 \theta_0 > 0.8$, $\theta_0$ being the neutrino mixing angle
determined by $V_l$
\footnote{This is the case if the angle is to lie in the first quadrant. 
As we shall see in 
detail in the next section, there can be nontrivial solutions for
$\theta_0$ in the second quadrant also.}.
This restricts one to
specific allowed ranges in the ratio of the vev's  $v'_\m$ and
$v'_\tau$, for each value of the ratio 
${\epsilon_2}/{\epsilon_3}$. Corresponding to these allowed values,
the 2-dimensional space
in the soft parameters $B_2$ and $B_3$ also gets constrained.
In the next section, we shall present
a detailed map of the region of the parameter space which corresponds
to large angle mixing as depicted by the SK data.

Before we end this section, we also list the three scalar mass matrices 
for this theory in basis {\it 1}. These include both the Higgs and the
slepton/sneutrino sectors, with the appropriate kinds of mixing between them.

The charged scalar mass matrix is given by

\begin{equation}
\textfont1=\scriptfont1 \scriptfont1=\scriptscriptfont1
\textfont0=\scriptfont0 \scriptfont0=\scriptscriptfont0
\textfont9=\scriptfont9 \scriptfont9=\scriptscriptfont9
M_c^2= \left( \begin{array}{cccccc}
s+\alpha_c^\prime+f_{3\tau}^2 &
             -B\mu+\alpha_{12} &
	      \mu \e_3+\alpha_{1\tau}- {f_{31}f_{3\tau} \over 2} &
	      \mu \e_2+\alpha_{1\mu}- {f_{21}f_{2\mu} \over 2}&
	         -\e_3f_{32}-Af_{3\tau} &
	         -\e_2f_{22}-Af_{2\mu} \\
             -B\mu+\alpha_{12} & r-\alpha_c^\prime &
             -B_3\e_3+ \alpha_{2\tau} &
	     -B_2\e_2+ \alpha_{2\mu} &
	    -\e_3 f_{31} &
	    -\e_2 f_{21} \\
	      \mu \e_3+\alpha_{1\tau}- {f_{31}f_{3\tau} \over 2} &
             -B_3\e_3+ \alpha_{2\tau} &
	     p_\tau + \alpha_{t_\tau} + \alpha'_c & \e_2 \e_3 & 
	     \m f_{32} + A f_{31} & 0 \\
	      \mu \e_2+\alpha_{1\mu}-{f_{21}f_{2\mu} \over 2}&
	     -B_2\e_2+ \alpha_{2\mu} & \e_2 \e_3 & p_\m +
	     \alpha_{t_\m} + \alpha'_c & 0 & \m f_{22} + A f_{21} \\  
	         -\e_3f_{32}-Af_{3\tau} &
	    -\e_3 f_{31} & \m f_{32} + A f_{31} & 0 
	    & q_\tau - 2 \alpha'_c + f^2_{3 \tau}
	    & f_{2\m} f_{3\tau} \\
	         -\e_2f_{22}-Af_{2\mu} &
	    -\e_2 f_{21} & 0 &
	    \m f_{22} + A f_{21} &
	    f_{2\m} f_{3\tau} & q_\m - 2\alpha'_c + f^2_{2\m}
	   \end{array} \right)
\end{equation}	   

with

$$r = m_{2}^2 + {\frac 1 4}{g^2}(v_{1}^2 + v_{2}^2 + v_{\tau}^2 +
v_{\m}^2)$$
$$s = m_{1}^2 + {\frac 1 4}{g^2}(v_{1}^2 + v_{2}^2 - v_{\tau}^2 -
v_{\m}^2 )$$
$$p_\tau = m_{{\tilde \tau}_L}^2 + f^2_{31}$$
$$q_\tau = m_{{\tilde \tau}_R}^2 + f^2_{31}$$
$$p_\m = m_{{\tilde \m}_L}^2 + f^2_{21}$$
$$q_\m = m_{{\tilde \m}_R}^2 + f^2_{21}$$
$$t_\tau = (-v_{1}^2 + v_{2}^2 + v_{\tau}^2 -v_{\m}^2)$$
$$t_\m = (-v_{1}^2 + v_{2}^2 - v_{\tau}^2 + v_{\m}^2)$$
\vskip 5pt
$${\frac {1} {4}} {g'}^2 c = \alpha'_c;~~~
{\frac {1} {2}} g^2 v_1 v_2  = \alpha_{12};~~~
{\frac {1} {2}} g^2 v_1 v_\tau  = \alpha_{1\tau};~~~ {\frac {1}{2}} 
g^2 v_2 v_\tau = \alpha_{2\tau}$$
$${\frac {1}{2}} g^2 v_2 v_\m = \alpha_{2\m};~~~ {\frac {1} {2}} g^2 v_1 
v_\m  = \alpha_{1\m};~~~
{\frac {1} {4}} g^2 t_\tau = \alpha_{t_\tau};~~~
{\frac {1} {4}} g^2 t_\m = \alpha_{t_\m}$$
$$f_3 v_\tau = f_{3 \tau};~~~ f_3 v_1 = f_{31};~~~ f_3 v_2 = f_{32}$$
$$f_2 v_\m = f_{2 \m};~~~ f_2 v_1 = f_{21};~~~ f_2 v_2 = f_{22}$$

\noindent
where both the left-and the right-chiral sleptons for each flavour have 
been included, the basis being
($H_1$, $H_2$, ${\tilde \tau}_L$, $\tilde{\mu}_L$, ${\tilde \tau}_R$, 
${\tilde \mu}_R$). Similarly, in the bases ($Re(H_1)$, $Re(H_2)$, $Re({\tilde
\nu}_\tau)$, $Re({\tilde \nu}_\mu$)) and ($Im(H_1)$, $Im(H_2)$,
$Im({\tilde \nu}_\tau)$, $Im({\tilde \nu}_\mu$))
respectively, the neutral scalar and the pseudoscalar mass matrices are 
given by
  
\begin{equation}
M_{s}^2 = \left( \begin{array}{cccc} 
m_{1}^2+2{\l}c+4{\l}v_{1}^2 & -4{\l}{v_1}{v_2}+{B}{\m} &
4{\l}{v_1}{v_\tau}+{\m}{\e_3} & 4{\l}{v_1}v_\mu + \mu \epsilon_2 \\
-4{\l}{v_1}{v_2}+{B}{\m} & m_{2}^2-2{\l}c+4{\l}v_{2}^2 &
-4{\l}{v_2}{v_\tau}+{B_3}{\e_3} & -4{\l}v_2 v_\mu + B_2 \epsilon_2 \\
4{\l}{v_1}{v_\tau}+{\m}{\e_3} &
-4{\l}{v_2}{v_\tau}+{B_3}{\e_3} & {m_{{\tilde {\n}}_{\tau}}^2}
+2{\l}c+4{\l}v_{\tau}^2 & \e_2 \e_3 + 2 {\l} v_\m v_{\tau} \\
4{\l}{v_1}{v_\mu}+{\m}{\e_2} &
-4{\l}{v_2}{v_\mu}+{B_2}{\e_2} & \e_2 \e_3 + 2 {\l} v_\mu v_\tau 
& {m_{{\tilde {\n}}_{\mu}}^2} +2{\l}c+4{\l}v_{\mu}^2
\end{array}  \right)
\end{equation}

\noindent
and
 
\begin{equation}
M_{p}^2 = \left( \begin{array}{cccc}
m_{1}^2+2{\l}c & -{B}{\m} &
{\m}{\e_3} & \m \e_2 \\
-{B}{\m} & m_{2}^2-2{\l}c &
-{B_3}{\e_3} & -B_2 \e_2 \\
{\m}{\e_3} &
-{B_3}{\e_3} & {m_{{\tilde {\n}}_{\tau}}^2}+2{\l}c & \e_2 \e_3 \\
{\m}{\e_2} &
-{B_2}{\e_2} & \e_2 \e_3 & {m_{{\tilde {\n}}_{\mu}}^2}+2{\l}c
\end{array}  \right)
\end{equation}

It may be remarked that the charged scalar and the neutral pseudoscalar mass
matrices will each have a zero eigenvalue, corresponding to the 
Goldstone bosons. Also, these scalar matrices have to be used for a complete
determination of the allowed space in $B_2$ and $B_3$, since a number of 
conditions related to electroweak symmetry breaking need to be satisfied before
values of these parameters, as extracted from  
Eqns. (8) and (9), pass off as valid ones.  These conditions
\cite{BOUND} include
the requirement that the potential be bounded from below, the negativity of
the vev's and non-negativity of the eigenvalues. They can be obtained
by a straightforward generalisation of the corresponding conditions
given in reference 18 where lepton number violation in only one family
has been considered.

\section{The constraints}

As we have already mentioned, our first constraint is on the
quantity $v'$. The allowed range of $\Delta m^2_{\mu\tau}$, combining the
fully contained events, partially contained events and upward-going muons,
is about $1.5 - 6.0 \times 10^{-3}$ eV$^2$ at 90\% confidence level 
\cite{SK2}. 
For the lightest 
neutralino mass varying between 50 and 200 GeV, this corresponds to
$v^{'} = 0.0001 - 0.0003~~$GeV (100--300 keV) to a fair degree of approximation.
Since $v^{'} = \sqrt{v^{'2}_{\mu} + v^{'2}_{\tau}}$, this automatically
puts a constraint on $v^{'}_{\mu}$ and $v^{'}_{\tau}$, and hence on
the vev's in basis {\it 1} for each value of the $\epsilon$-parameters. 

Next, the value of the angle $\theta$ should determine $v^{'}_{\mu}$ and 
$v^{'}_{\tau}$ completely, once $v^{'}$ is fixed.
As we have pointed out in the previous section, for each value of
$\epsilon_2/\epsilon_3$, one is restricted to particular values of 
$\theta$ in order that the condition of large angle mixing, namely
$\sin^2 2{\theta}_0 > 0.8$, is satified in terms of the ultimate
neutrino mixing angle $\theta_0$. In Figs. 1 and 2 we outline these
allowed regions in the parameter space of $\epsilon_2/\epsilon_3$ vs. $\theta$,
for   $\epsilon_2/\epsilon_3 \le 1$ and  $\epsilon_2/\epsilon_3 \ge 1$
respectively. The allowed bands are found to be sensitive to the ratio
rather than the actual values of the $\epsilon$'s. The
interesting point to note here is that there are two values
of $\theta$ for each value of the ratio along the x-axis. This is due to 
the two
solutions with angles in either of the first or the second quadrant,
both cases yielding the same value for the oscillation probability.
Physically, the second case corresponds to the superposition
of the $\nu_\mu$ and $\nu_\tau$ states  being performed with an extra 
phase rotation through an angle $\pi$ for one of them. In other words,
the two solutions represent situations with the two neutrino flavours having
identical and opposite CP-properties respectively. 

The angles $\theta$ and $\theta_0$ are practically equal if 
$\frac{\epsilon_2}{\epsilon_3} << \frac{m_\mu}{m_\tau}$. 
In this case the
off-diagonal elements of the charged lepton mass matrix ${\cal{M}}_l$ 
in basis {\it 2} are neligibly small compared to the diagonal ones. However,
Figs. 1 and 2 clearly demonstrate that such a large hierarchy between the
two bilinear terms in the superpotential is by no means a necessity 
and that in fact it is possible to have them not only with the same order 
of magnitude but also with an inverted hierachy as well.

Each point in the allowed regions shown in Figs. 1 and 2 corresponds
to a theory for which the two bilinear soft terms
$B_2$ and $B_3$ can be calculated. The ranges of values thus obtained have
been plotted in Figs. 3 and 4, again with opposite hierarchies of
the parameters $\epsilon_2$ and $\epsilon_3$. The two bands in 
each figure arise from the two solutions of $\theta$ for
each ratio of the $\epsilon$'s and a particular value of $\epsilon_3$ 
(Fig. 1) or $\epsilon_2$ (Fig. 2). This underlines the fact that the two 
solutions for the angle in the two quadrants are indeed physically distinct, 
with all the different phenomenological implications of the two combinations
of the $\mathcal L$-violating soft terms \footnote{In fact, there could be two more
sets of solutions in the two remaining quadrants. However, these solutions
are not distinct from the previous ones, as they can be
mapped back to the cases of same and opposite parity for the two neutrinos.
Explicit computaions also lead to the two already obtained combinations of
$B_2$ and $B_3$.}.

The value of $B_2$ in Fig. 3 
is found to become very large when $\epsilon_2$ is assigned a 
value much smaller than that of $\epsilon_3$ (and similarly for $B_3$ in 
Fig. 4
with a very small $\epsilon_3$). This is analogous to the $\mu$-B problem
of the MSSM. However, Figs. 1 and 2 already tell us that such a 
large hierarchy of the two $\epsilon$'s is {\em not} a necessary condition 
for large angle neutrino mixing. The very expressions for the 
$B$-parameters show 
that except in cases where very small $\epsilon_{2(3)}$ jacks  
them up to large values, they are also controlled by the scale of the slepton 
mass. We also see from Fig. 3 that  $B_3$ cannot attain values 
much larger than the electroweak scale so long as $\epsilon_2 \le \epsilon_3$.
The same feature is observed for $B_2$ in Fig. 4. 

However, it must be noted at this point that the sets of values for 
$B_2$
and $B_3$ as seen in Fig. 3 or 4 result from the choices of two 
parameters, {\it viz.}, $\epsilon_3$ (Fig.3) or $\epsilon_2$ (Fig.4) and 
their appropriate ratios, the
choices being rather specific and spanned over identical regions in these two  
figures. This is the 
reason why one
of $B_2$ and $B_3$ ranges over several orders of magnitude while the other is
highly constrained in a particular figure. Here, it should  be made clear 
that the magnitudes of $B_2$ and $B_3$ are mainly determined by those 
of $\epsilon_2$ and $\epsilon_3$ respectively for very small values 
($< 10^{-5}$) of the latter two. As, neither the smallness of the individual 
values of $\epsilon$'s nor that of their ratios is anyway restricted, 
we can always tune these two parameters so that both of $B_2$ (Eqn. (8)) 
and $B_3$ (Eqn. (9)) are 
simultaneously very large (compared to the electroweak scale) leading to 
additional regions in the allowed 
$B_2-B_3$ space which are absent in Figs. 3 and 4. 

%
%

Summing all these up,
the parameter space of the soft terms (modulo their additional dependence on 
the MSSM parameters) is seen to be sufficiently free from any requirement of
fine-tuning or large hierarchy, and large angle neutrino mixing 
as well as the expected neutrino mass hierarchy in the second and third 
families is reproduced over a rather wide range in the space of
R-parity violating parameters. Also, $B_2$ and $B_3$ are not compelled to
show any hierarchical behaviours with respect to the R-conserving
soft term $B$, and any approach connecting them at a high scale is
consistent with the range of values allowed here.

The discussion becomes more transparent if we break up the results shown
in  Figs. 3 and 4 into 3 broad regions, viz., (i) very small
$\epsilon_2(\epsilon_3)$ ($\sim 10^{-8}$), (ii) somewhat intermediate values
of $\epsilon_2(\epsilon_3)$ ($\sim 10^{-5}$) and (iii) rather large values of
$\epsilon_2(\epsilon_3)$ ($\sim 10^{-2}$) in reference to Eqn. (8) (Eqn. (9)).
The numbers in the parenthesis
correspond to the left extreme graphs of Fig. 3 (Fig. 4).
In case (i) $\epsilon_2 (\epsilon_3)$ dependent terms lead and its presence
in the denominators with a very small value leads to a large value for
$B_2(B_3)$.  When $\epsilon_2(\epsilon_3)$ increases as for case (ii) the
contributions from these terms
gradually become comparable to that coming from the 4th term which
depends only upon MSSM parametrs, viz., $\mu$ and $\tan \beta$. This
restricts $B_2
(B_3)$ to intermediate values. The relative sign between the 4th term and other
terms collectively is instrumental in fixing the sign of $B_2(B_3)$ in this
region and is clearly visible as we proceed along the second row of graphs in
Fig. 3 (Fig. 4). In case (iii) the 4-th term controls
$B_2(B_3)$. Naturally, in this case, the order of $B_2(B_3)$ is set by
$\mu$ and
$\tan \beta$ and is restricted to be around 100 GeV for choices of
these two parameters that are compatible with existing collider data.

In the results presented above, $\mu$ and $\mu^{'}$ have been taken to be of
the same sign, with both $\epsilon_2$ and $\epsilon_3$ having signs opposite to
it. On reversing this relative sign (and also that between the two 
$\epsilon$-terms), it is seen that the orders of
magnitude of $B_2$ and $B_3$ do not change; however, the signs of one or 
both of them are liable to get reversed. 

On reversing the sign of $\mu$,
both $B_2$ and $B_3$ may change sign if the bulk 
contributions to them come from the 4-th terms in Eqns. (8) and (9). 
Otherwise, the relative sign between $B_2$ and $B_3$ and also their
magnitudes 
(to a good approximation) are preserved under such reversal. Changing the 
sign of $\mu$ should be accompanied by a change in sign of $B$ (of MSSM) 
to retain a relative sign between themselves which is a must to achieve 
electroweak symmetry breaking at the proper scale. On the 
other hand, $B_2$ and $B_3$ pick up a relative sign on allowing for the 
same between $\epsilon_2$ and $\epsilon_3$ only if the latter two 
contribute heavily to $B_2$ and $B_3$ respectively.

We restrict ourselves by presenting only two sets 
of figures (Figs. 3  and 
4) which jointly cover rather wide intervals of $\epsilon_2$ and
$\epsilon_3$. Results of any combinatorics of relative signs between 
$\mu$ and  the $\epsilon$'s can be directly estimated to a fair degree of
accuracy from these two sets of figures. These show that the magnitudes of
$B_2$ and $B_3$ can range over several orders of magnitude with any
relative sign between them.

It should be re-iterated that variations of $B_2$ ($B_3$) with the
MSSM parameters $\mu$ and $\tan \beta$ are likely to be most pronounced when
major contributions to it comes from the 4th term in Eqn. (8) (Eqn. (9)).
This should be contrasted with the results presented in Figs. 1 and 2
where the allowed band is quite insensitive to the MSSM parameters.
$B_2$ and $B_3$ exhibits such a sensitivity  when $\epsilon_2$
($\epsilon_3$) is $\simeq 10^{-3}$ or more, so that the other terms in
Eqn. (8) (Eqn. (9)) have 
comparable orders of magnitude ($\simeq$ electroweak scale) leading to
$B_2$ ($B_3$) lying around
100 GeV. Here, variation of $\tan \beta$ over a range 5 to 50 can change
$B_2$ ($B_3$) at most by 1 order while variation with $\mu$ over a range
$-$500 to $+$500 does not change things much, although a flip in the relative
sign between $B_2$ and $B_3$ may be a possible outcome of the variation
of $\mu$.

In the above, we have not discussed the mass-splitting
required between the electron and the muon neutrinos for explaining the
solar neutrino problem. As we have already mentioned, the 
trilinear interactions in Eqn. (2) give rise
to loop-induced masses that can account for this splitting rather well.
The nature of the splitting will depend on whether one wants to
provide an explanation in terms of matter-enhanced or vacuum oscillations,
and the only consequent constraints will come on the respective
$\lambda$-and $\lambda^{'}$-couplings in basis {\it 2}.

There are, however, additional issues that must be addressed here. 
In general, the very structure of the charged scalar, neutral scalar 
and pseudoscalar mass terms, obtainable from Eqn. (3),
admits of mixing between the smuon and stau flavour states, as
also between the corresponding sneutrino flavours in both the
neutral scalar and pseudoscalar mass matrices. This in general causes
a mismatch between these scalar mass matrices and the charged lepton
mass matrix, standing in the way of their simultaneous diagonalisation.
This in general can lead to an enhancement of FCNC, 
contributing to processes like $\tau \longrightarrow \mu \gamma$.

FCNC suppression can in general be ensured by one of three mechanisms--
degeneracy, alignment and decoupling \cite{FCNC}. Decoupling in this 
case requires 
the slepton and sneutrino masses to be very high so that contributions to
the relevant processes from loop diagrams can be suppressed. With these
masses on the order of 100 GeV, one has to ensure either alignment of
the scalar mass matrices with the one for charged leptons, or a close
degeneracy of the scalar eigenvalues. Both conditions are seen to
be satisfied in this case if $({M^2_{\tilde{L}_2 \tilde{L}_3}})_i << 
({M^2_{\tilde{L}_2 \tilde{L}_2}})_i$ 
for $i = s, p~~or~~c$ (i.e. in the neutral scalar, pseudoscalar or charged
scalar mass squared matrices).

Figures 5 and 6 contain plots of the ratio of the $\mu_L\tau_L$ and  
$\mu_L\mu_L$ elements for each of the three matrices, against the ratio of 
$\epsilon_2$ and  $\epsilon_3$ (again for both $\epsilon_2 \le \epsilon_3$ and  
$\epsilon_2 \ge \epsilon_3$). A pattern  is revealed in each case on
imposing the requirement on the off-diagonal elements to be small
compared to the diagonal ones.

First consider the case  $\epsilon_2 \le \epsilon_3$. For  small values
of this  ratio, the smallness of all the diagonal terms (and therefore
the suppression of FCNC) is assured, irrespective of the value of
$\epsilon_3$. In such cases it is enough to consider
the scalar mass matrices in basis {\it 2} itself, since the condition
$\epsilon_2 << \epsilon_3$ makes the charged lepton mass matrix 
practically diagonal in this basis.
The off-diagonal terms in this basis are controlled by
the quantity $v^{'2}$ which is constrained from neutrino masses to be 
much smaller than $m^2_{\tilde{l}}$ and $\mu^2$ which dominate the values
of the diagonal terms. However, as one increases $\epsilon_2 / \epsilon_3$,
the charged lepton mass matrix ceases to be diagonal in basis {\it 2}, and
the scalar mass matrices need to be evaluated {\em after} diagonalising it.
In such cases the off-diagonal elements are also influenced by the actual 
value of $\epsilon^2_3$. As can be seen from Fig. 5, for
$\epsilon_3$ approaching 100 GeV, the off-diagonal term 
starts becoming comparable to the diagonal ones
as the ratio  $\epsilon_2 / \epsilon_3$ is close to unity. Exactly
the same thing can be said of $\epsilon_2$ in Fig. 6 where predictions 
for an inverted ratio between the two $\epsilon$'s are shown.

A rather interesting conclusion follows from the above discussion. While a
compatibility with the SK results is generally observed over a large 
area in the space of R-parity violating parameters, the need to suppress
FCNC effects introduces at least one type of hierarchy. This is apparent from 
the fact that one of the two bilinear parameters  $\epsilon_2$ and
$\epsilon_3$  can be allowed to be as large as in the electroweak scale
if their mutual ratio has a hierarchy of 2 to 3 orders of magnitude.
On the other hand, allowing the two of them to be of the same order is
consistent with the suppression of FCNC only if the scale of their magnitude 
is small compared with the corresponding R-conserving parameter i.e.
$\epsilon_{2(3)}  <<  \mu$. Thus, either a hierarchy in the two R-violating
parameters in the superpotential or a smallness of the R-violating ones
compared to R-conserving ones is a necessary consequence of this
scenario.

\section{Conclusion}

We have investigated a SUSY theory where the bilinear R-parity violating terms
are responsible for tree-level neutrino mass and large angle mixing,
leading to a pattern that explains the observed defficiency of
atmospheric muonic neutrinos. The scenario also admits of trilinear R-parity
violating terms that cause the mass splitting necesary to explain the
solar neutrino puzzle. 

A detailed study of the parameter space of the theory shows that,
as far as constraints arising from the mass-splitting and large angle mixing 
are concerned, neither a hierarchy of R-violating parameters nor a 
fine-tuning among them is necassary, and a large area of the parameter
space of the R-breaking soft terms is `naturally' allowed. However, the 
suppression of FCNC can  require either the R-violating parameters in the
superpotential to be small compared to the $\mu$-parameter, or such
parameters themselves to have a mutual hierarchy of values. 

\vskip 5pt

{\bf Acknowledgment:} We acknowledge helpful discussions with A.
Joshipura, S. Vempati and F. Vissani. Our thanks are also due to 
R. Shrivastav and P.K. Mohanty for computational and graphics-related
help.

\newpage
\begin{figure}[htb]
\vspace*{-2.2in}
\centerline{\epsfig{file=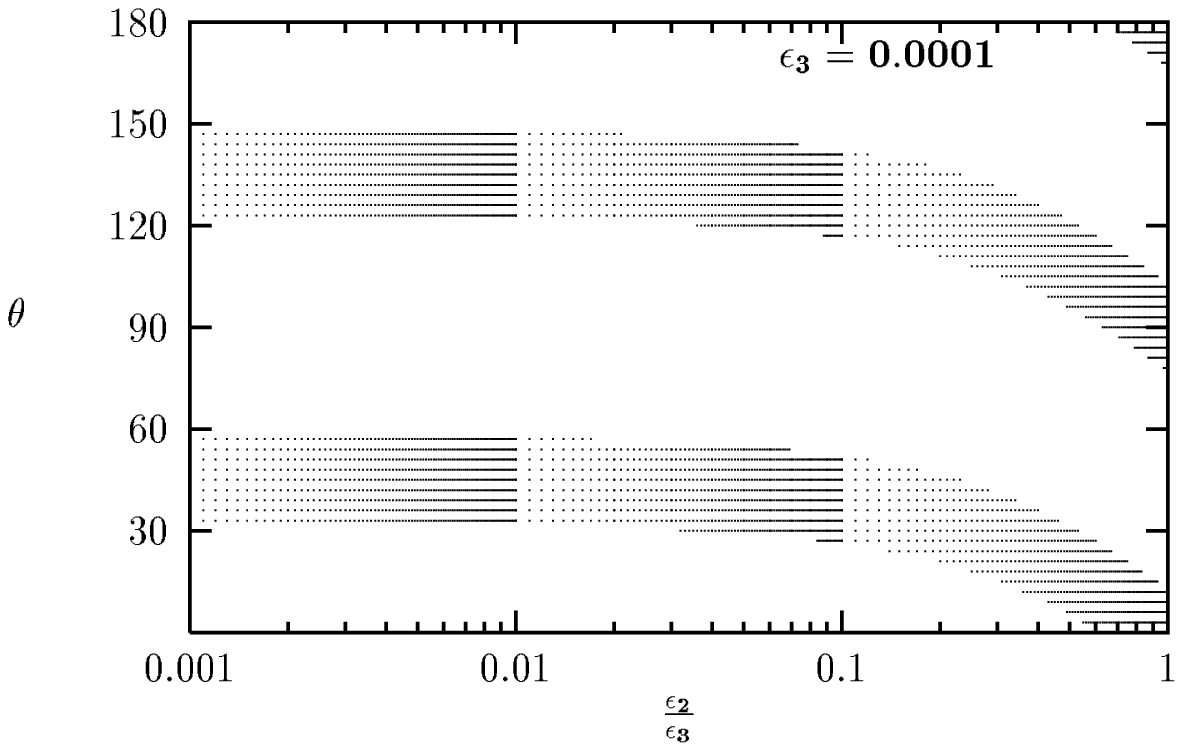,width=18cm}}
\vspace*{-5.9in}
\caption{Allowed bands of neutrino mixing angle ($\theta$ in degrees) 
corresponding
to maximal mixing between 2nd and 3rd generation of neutrinos are plotted 
against the ratio
of epsilons ($\frac{\e_2}{\e_3}$). The MSSM parameters are $\mu= -500$, 
$\tan \beta =10$ and $B=100$. The pattern remains almost the same for different
$\epsilon_3$'s. The different shadings arise due to varying number of sample
points in different regions.}
\end{figure}

\vskip 100pt
\begin{figure}[htb]
\vspace*{-2.4in}
\centerline{\epsfig{file=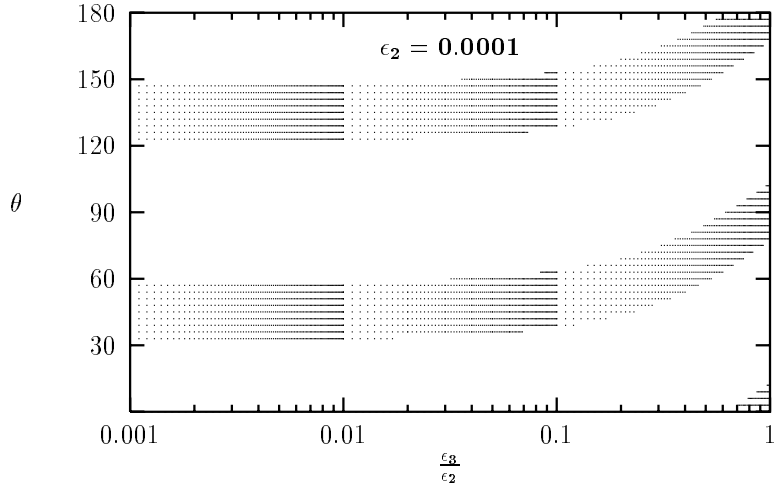,width=18cm}}
\vspace*{-5.9in}
\caption{Allowed bands of neutrino mixing angle ($\theta$ in degrees) 
corresponding
to maximal mixing between 2nd and 3rd generation of neutrinos is plotted 
against the ratio
of epsilons ($\frac{\e_3}{\e_2}$). The MSSM parameters are same as 
in Fig. 1. The pattern remains almost the same for different
$\epsilon_2$'s. The different shadings arise due to varying number of sample
points in different regions.}
\end{figure}

\begin{figure}[htb]
\centerline{\epsfig{file=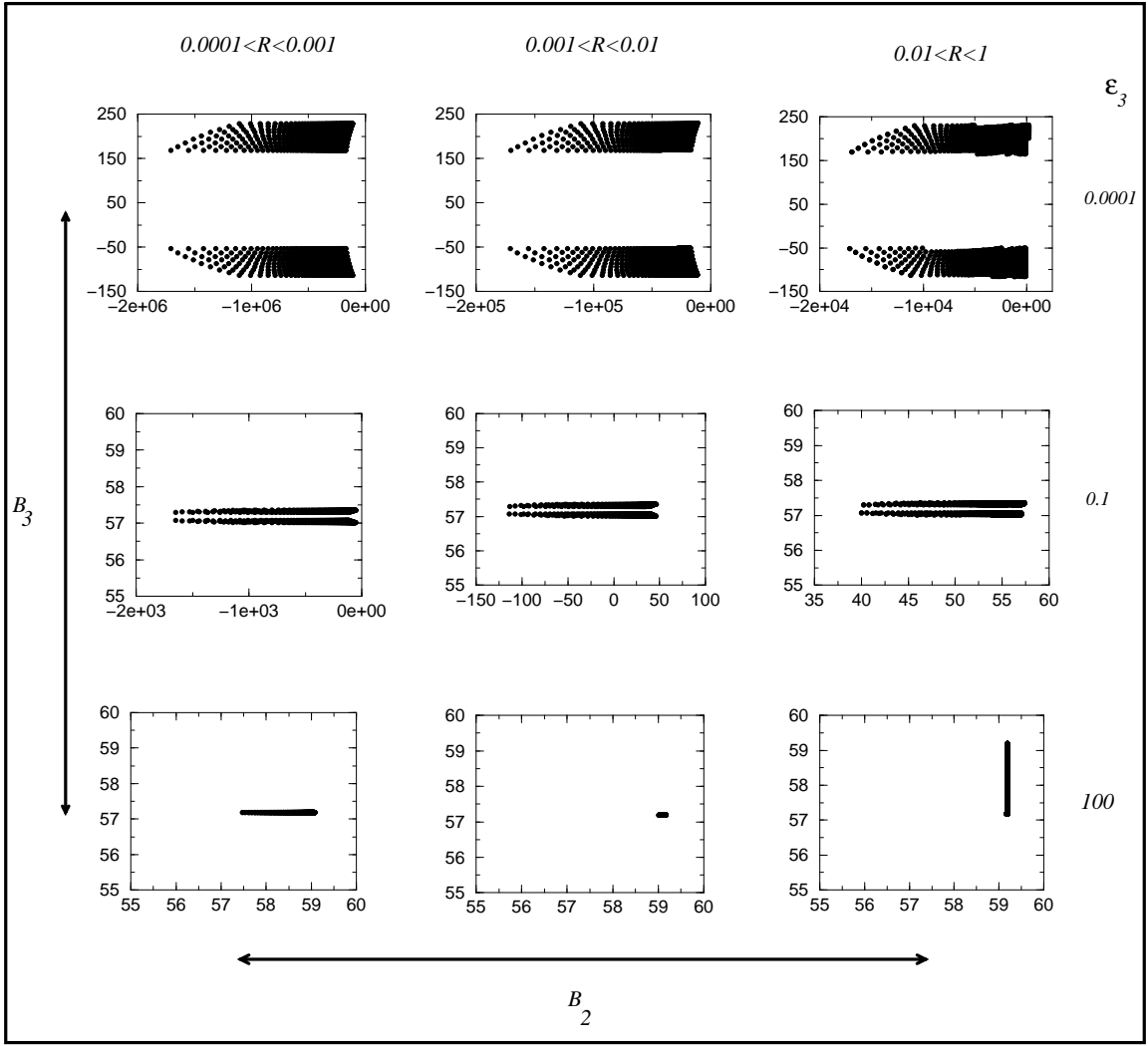,width=17cm}}
\vspace*{-2.0in}
\caption{Allowed regions of $B_2$ - $B_3$ space (in GeV). 
$R={\epsilon_2 \over \epsilon_3}$ varies along the columns as indicated in the 
top margin. $\epsilon_3$ varies along the rows as shown in the right margin.
The MSSM parameters are $\mu= -500$, $\tan \beta=10$ and the common
slepton mass at the electroweak scale is 200 GeV.} 
\end{figure}

\begin{figure}[htb]
\centerline{\epsfig{file=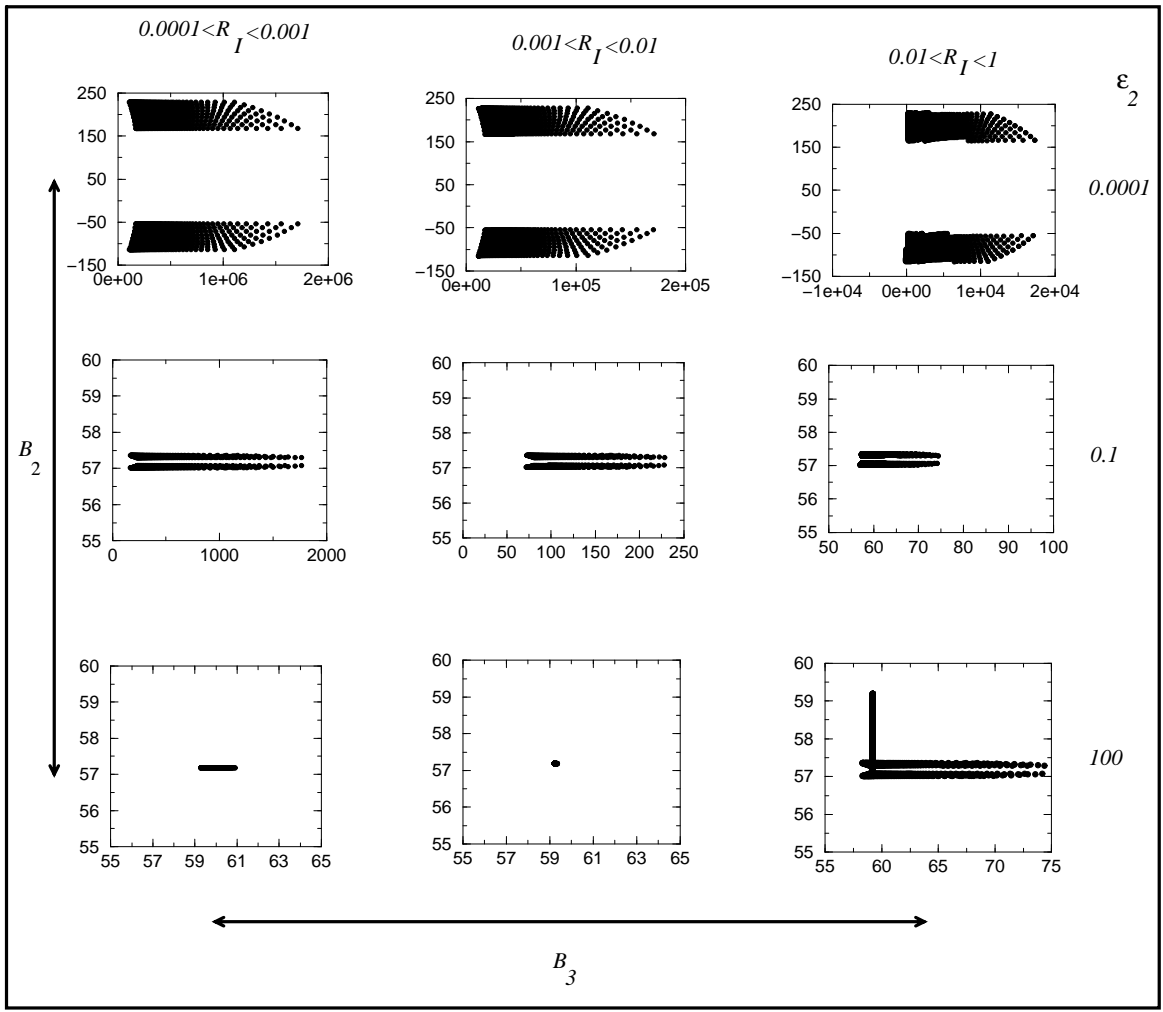,width=17cm}}
\vspace*{-2.0in}
\caption{Allowed regions of $B_2$ - $B_3$ space (in GeV). 
$R_I={\epsilon_3 \over \epsilon_2}$ varies along the columns as indicated
in the 
top margin. $\epsilon_2$ varies along the rows as shown in the right margin.
The MSSM parameters are same as in Fig. 3.}
\end{figure}

\begin{figure}[htb]
\vspace*{-2.2in}
\centerline{\epsfig{file=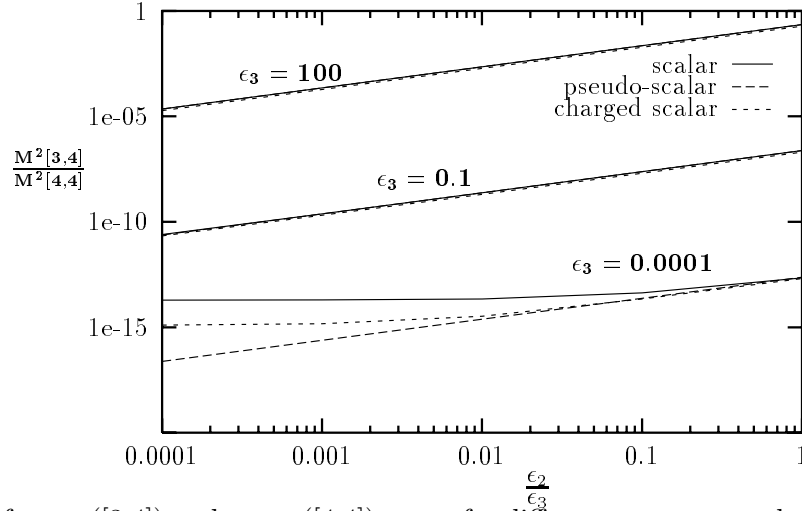,width=18cm}}
\vspace*{-6.1in}
\caption{Ratio of $\mu_L\tau_L$ ([3,4]) and $\mu_L\mu_L$ ([4,4]) terms for 
different
mass-squared matrices (in a basis where the lepton mass matrix is diagonal) 
is  plotted against the ratio
($\frac{\e_2}{\e_3}$). The MSSM parameters are $\mu= -500$,
$\tan \beta=10$, $B=100$ and common slepton mass at the electroweak scale is
200 GeV.}
\end{figure}

\begin{figure}[htb]
\vspace*{-1.4in}
\centerline{\epsfig{file=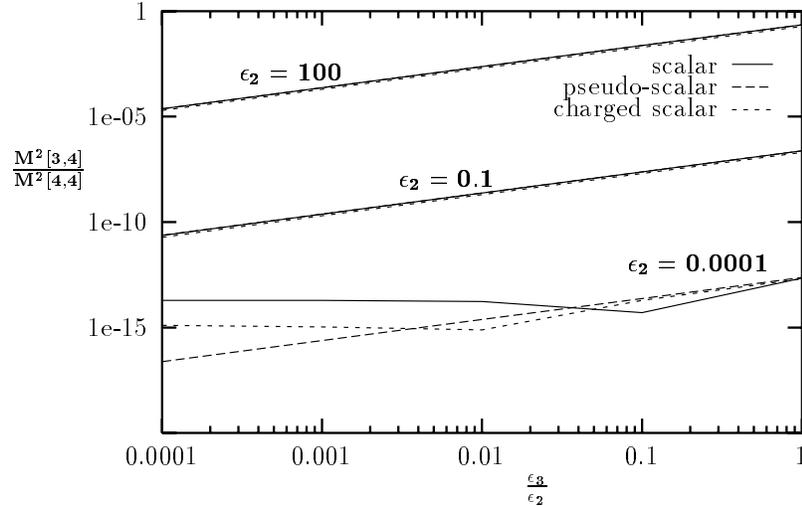,width=18cm}}
\vspace*{-6.1in}
\caption{Ratio of $\mu_L\tau_L$ ([3,4]) and $\mu_L\mu_L$ ([4,4]) terms for 
different mass-squared matrices (in a basis where the lepton mass matrix is 
diagonal) is plotted against the ratio
($\frac{\e_3}{\e_2}$). The MSSM parameters are same
as in Fig. 5.}
\end{figure}

\end{document}